\documentclass[11pt]{article}
\usepackage{Blois,epsfig}

\def\be{\begin{equation}}
\def\ee{\end{equation}}
\def\bea{\begin{eqnarray}}
\def\eea{\end{eqnarray}}

\begin{document}
\vspace*{2cm}
\begin{center}
\Large{\textbf{XIth International Conference on\\ Elastic and Diffractive Scattering\\ Ch\^{a}teau de Blois, France, May 15 - 20, 2005}}
\end{center}

\vspace*{2cm}
\title{DIFFERENT ASPECTS OF BFKL AT NLL}

\author{AGUST{\' I}N SABIO VERA}

\address{II. Institut f{\"u}r Theoretische Physik, Universit{\"a}t Hamburg\\Luruper Chaussee 149, 22761 Hamburg, Germany}

\maketitle\abstracts{Progress in the understanding of the BFKL 
approach in the NLL approximation is reported. The study based on 
the iteration of the kernel using the exponentiation of the gluon Regge 
trajectory is reviewed in QCD and N=4 super Yang--Mills theories. 
Properties of the gluon Green's function in the high energy Regge limit 
for forward and non--forward scattering are considered. A 
novel representation of collinearly improved kernels is also presented.}

\section{The High Energy behaviour of QCD in the Regge limit}

A very challenging aspect of QCD which remains to be understood is the behaviour 
of scattering amplitudes when the center--of--mass energy is much larger than any 
other scale. In this limit the Balitsky--Fadin--Kuraev--Lipatov (BFKL) 
approach~\cite{FKL}, based on the all--orders resummation of logarithms in energy, 
provides a very useful tool to handle different scattering processes. 

This contribution is based on the analysis of the BFKL kernel and 
gluon Green's function (GGF) at next--to--leading (NLL) order 
where $(\alpha_s \ln{s})^n$  
and $\alpha_s (\alpha_s \ln{s})^n$ terms are considered~\cite{FLCC}. 
This accuracy is needed to understand the r{\^o}le of the 
running coupling and to fix the energy scale in the logarithms. Many studies have been devoted to the analysis of the NLL GGF  
({\it e.g.}~\cite{Salam,Ciafalonietal,NLOpapers}). Here a framework suitable to extract the GGF from the NLL BFKL integral equation is 
explained in some detail. In \cite{Andersen:2003an} it was shown how to remove 
poles in $4 + 2 \epsilon$ dimensional regularisation by introducing  
a logarithmic dependence on a mass parameter $\lambda$ without angular averaging 
the NLL kernel. In this 
regularisation it is then useful to iterate the BFKL equation for the $t$--channel 
partial wave generating poles in the $\omega$--plane. Performing the Mellin transform back to energy space the NLL GGF reads
\begin{eqnarray}
\label{solution1}
f({\bf k}_a ,{\bf k}_b, {\rm Y}) &=& 
e^{\omega_0^\lambda \left({\bf k}_a\right) {\rm Y}} \frac{}{}\delta^{(2)} ({\bf k}_a - {\bf k}_b)\nonumber\\
&+&\sum_{n=1}^{\infty} \prod_{i=1}^{n} 
\int d^2 {\bf k}_i \int_0^{y_{i-1}} d y_i \left[\frac{\theta\left({\bf k}_i^2 - \lambda^2\right)}{\pi {\bf k}_i^2} \xi\left({\bf k}_i\right) +\widetilde{\mathcal{K}}_r \left({\bf k}_a+\sum_{l=0}^{i-1}{\bf k}_l,
{\bf k}_a+\sum_{l=1}^{i}{\bf k}_l\right)\frac{}{}\right]\nonumber\\
&  \times&  e^{
\omega_0^\lambda\left({\bf k}_a+\sum_{l=1}^{i-1} {\bf k}_l
\right)(y_{i-1}-y_i)}\  e^{\omega_0^\lambda\left({\bf
  k}_a+\sum_{l=1}^i {\bf k}_l\right)y_n}\delta^{(2)} \left(\sum_{l=1}^{n}{\bf
k}_l  
+ {\bf k}_a - {\bf k}_b \right),
\end{eqnarray}
where the first integral in rapidity has an upper limit $y_0 = {\rm Y}$. 
The initial term in the expansion corresponds to two Reggeized gluons propagating 
in the $t$--channel. The dependence on the gluon Regge 
trajectory
\begin{eqnarray}
\label{trajectory}
\omega_0^\lambda \left({\bf q}\right) &=& 
- \bar{\alpha}_s \ln{\frac{{\bf q}^2}{\lambda^2}}
+ \frac{\bar{\alpha}_s^2}{4}\left[\frac{\beta_0}{2 N_c}\ln{\frac{{\bf q}^2}{\lambda^2}}\ln{\frac{{\bf q}^2 \lambda^2}{\mu^4}}+\left(\frac{\pi^2}{3}-\frac{4}{3}-\frac{5}{3}\frac{\beta_0}{N_c}\right)\ln{\frac{{\bf q}^2}{\lambda^2}}+6 \zeta(3)\right],
\end{eqnarray}
exponentiates, corresponding to no--emission probabilities between two consecutive 
effective vertices. The real emission consists of two parts:
\begin{eqnarray}
\xi \left({\rm X}\right) &\equiv& \bar{\alpha}_s +  
\frac{{\bar{\alpha}_s}^2}{4}\left(\frac{4}{3}-\frac{\pi^2}{3}+\frac{5}{3}\frac{\beta_0}{N_c}-\frac{\beta_0}{N_c}\ln{\frac{{\rm X}}{\mu^2}}\right),  
\end{eqnarray}
which cancels the singularities present in the trajectory order by order in 
${\bar\alpha}_s$, and $\tilde{\cal K}_r$, which does not generate singularities 
when integrated over the emissions' phase space.

\begin{figure}
\centering
\vspace{5cm}
\includegraphics{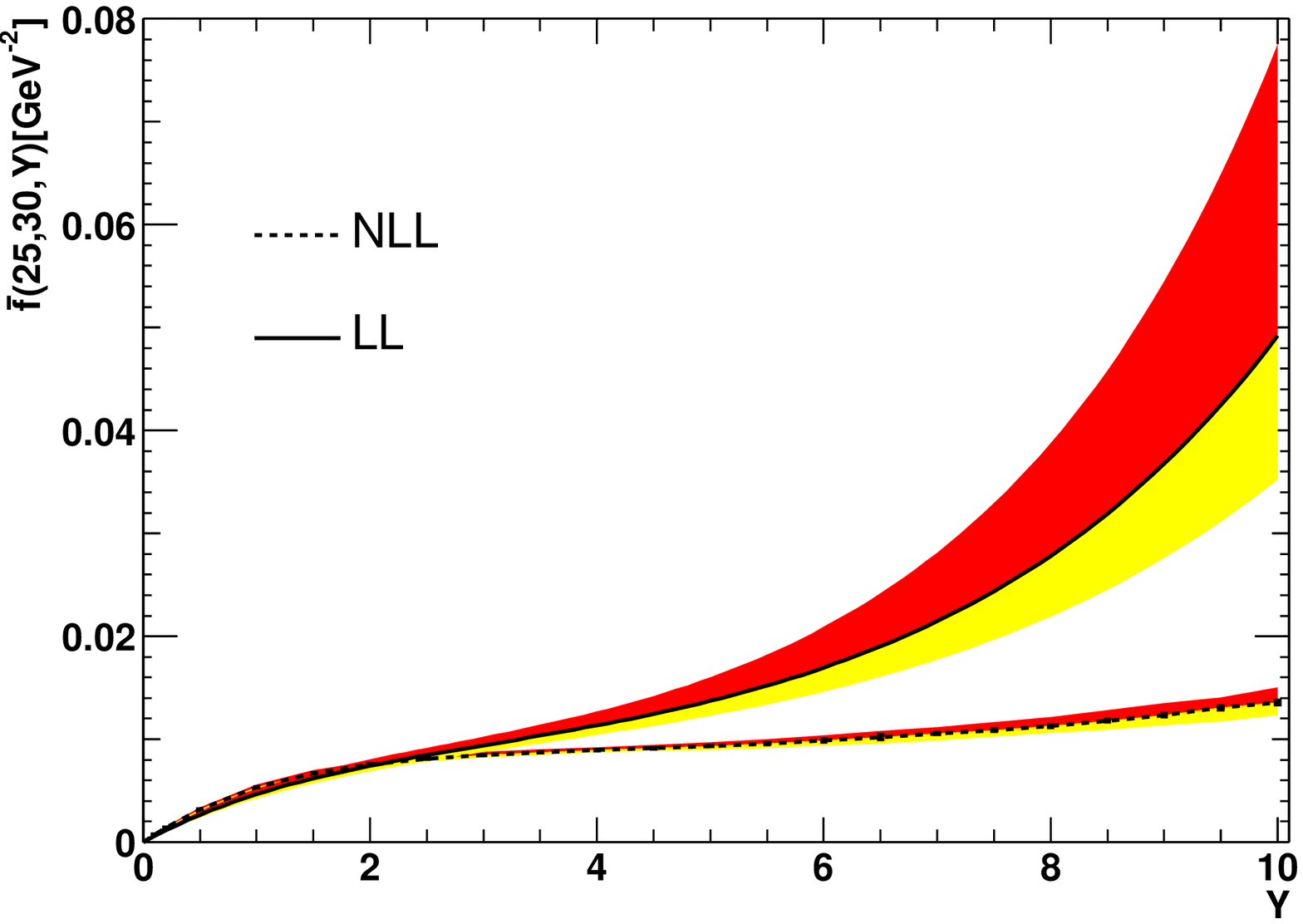}
\hspace{0.2cm}
\includegraphics{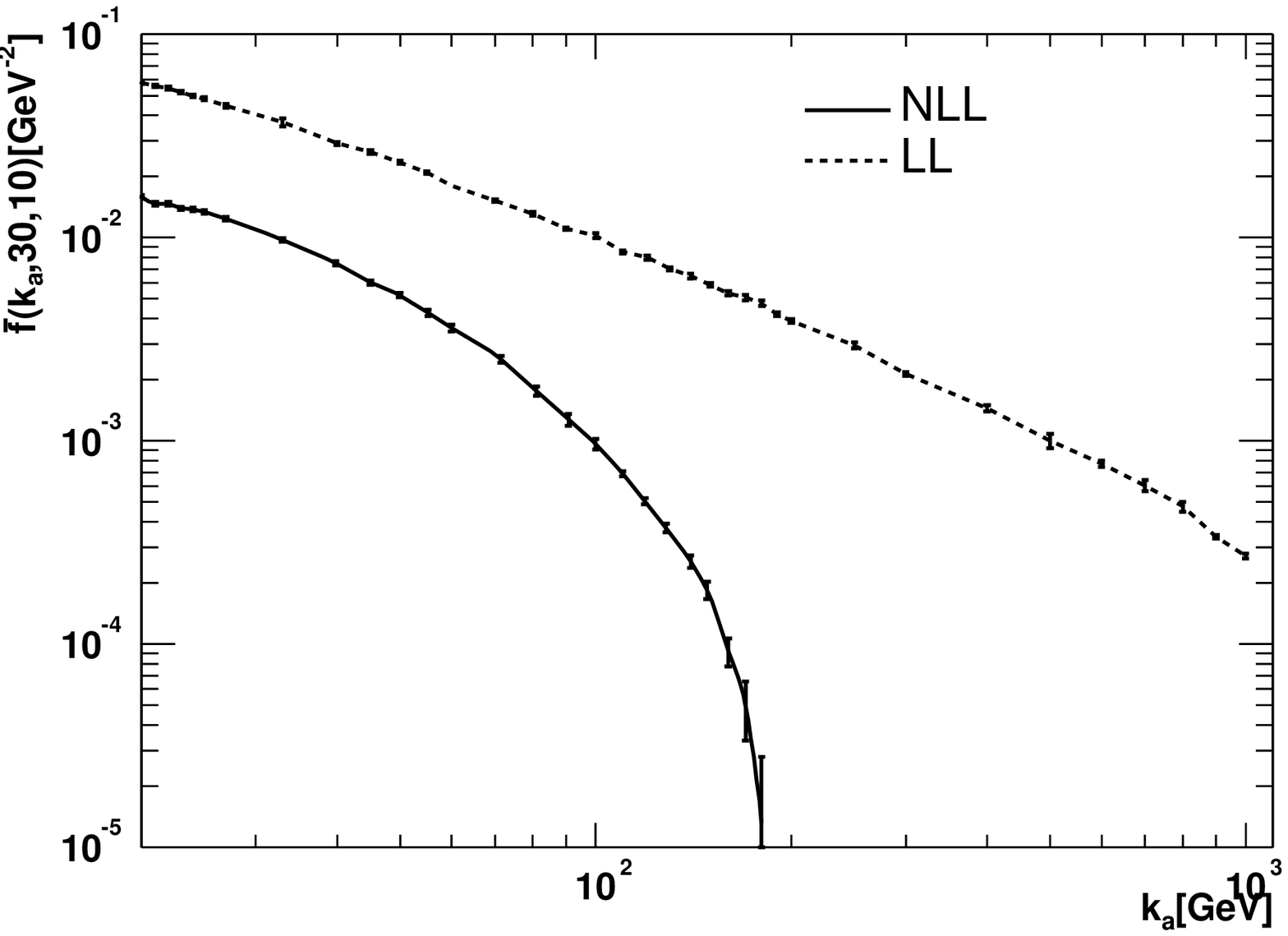}
\caption{The NLL GGF: Evolution in rapidity at LL and NLL for $k_a=25$~GeV and $k_b=30$~GeV (left), and the dependence on $k_a$ for fixed $k_b=30$~GeV and $Y=10$ (right).}
\label{JAfigure1}
\end{figure}

The numerical analysis of the NLL GGF was performed in~\cite{Andersen:2003wy}. 
As expected the intercept is lower than at leading--logarithmic (LL) 
accuracy. This is shown at the left hand side of Fig.~\ref{JAfigure1} where the 
coloured bands correspond to different choices of renormalisation scale.
In $k_t$--space the NLL corrections are stable when the two
transverse scales entering the forward GGF are similar. If they are very different 
then the convergence is not good, having an oscillatory behaviour with regions of 
negative values along the period of oscillation (see second plot of 
Fig.~\ref{JAfigure1}).

It is possible to improve the convergence of the expansion~\cite{Salam,Ciafalonietal,NLOpapers}. An original approach was suggested in~\cite{Salam} based on the 
introduction of an all--orders resummation of terms compatible with 
renormalisation group evolution. In~\cite{Vera:2005jt} it has been shown how to 
apply this $RG$--resummation to the iterative solution here described. In a nutshell: For small $\bar{\alpha}_s$ the solution to the $\omega$--shift in~\cite{Salam} 
\begin{eqnarray}
\label{shiftGS}
\omega &=& {\bar \alpha}_s \left(1+\left({\rm a}+\frac{\pi^2}{6}\right){\bar \alpha}_s\right) \left(2 \psi(1)-\psi\left(\gamma+\frac{\omega}{2}-{\rm b}\,{\bar \alpha}_s \right)-\psi\left(1-\gamma+\frac{\omega}{2}-{\rm b}\,{\bar \alpha}_s \right)\right)\nonumber\\
&+& {\bar \alpha}_s^2 \left(\chi_1 \left(\gamma\right) 
+\left(\frac{1}{2}\chi_0\left(\gamma\right)-{\rm b}\right)\left(\psi'(\gamma)+\psi'(1-\gamma)\right)-\left({\rm a}+\frac{\pi^2}{6}\right)\chi_0(\gamma)\right),
\end{eqnarray}
can be approximated by the sum of the solutions to the shift at each of the poles 
of the LL eigenvalue of the kernel, {\it i.e.}
\begin{eqnarray}
\label{All-poles}
\omega &=& \bar{\alpha}_s \chi_0 (\gamma) + \bar{\alpha}_s^2 \chi_1 (\gamma) 
+ \left\{\sum_{m=0}^{\infty} \left[\left(\sum_{n=0}^{\infty}
\frac{(-1)^n (2n)!}{2^n n! (n+1)!}\frac{\left({\bar \alpha}_s+ {\rm a} \,{\bar \alpha}_s^2\right)^{n+1}}{\left(\gamma + m - {\rm b} \,{\bar \alpha}_s\right)^{2n+1}}\right) \right. \right. \nonumber\\
&-&\left.\left.\frac{\bar{\alpha}_s}{\gamma + m} - \bar{\alpha}_s^2 \left(\frac{\rm a}{\gamma +m} + \frac{\rm b}{(\gamma + m)^2}-\frac{1}{2(\gamma+m)^3}\right)\right]+ \left\{\gamma \rightarrow 1-\gamma\right\}\right\}, 
\end{eqnarray}
where $\chi_0$ and $\chi_1$ are the LL and NLL scale invariant parts of the kernel,
 and a and b the coefficients of the single and double poles in the collinear 
limit.

\begin{figure}
\vspace{4cm}
\includegraphics{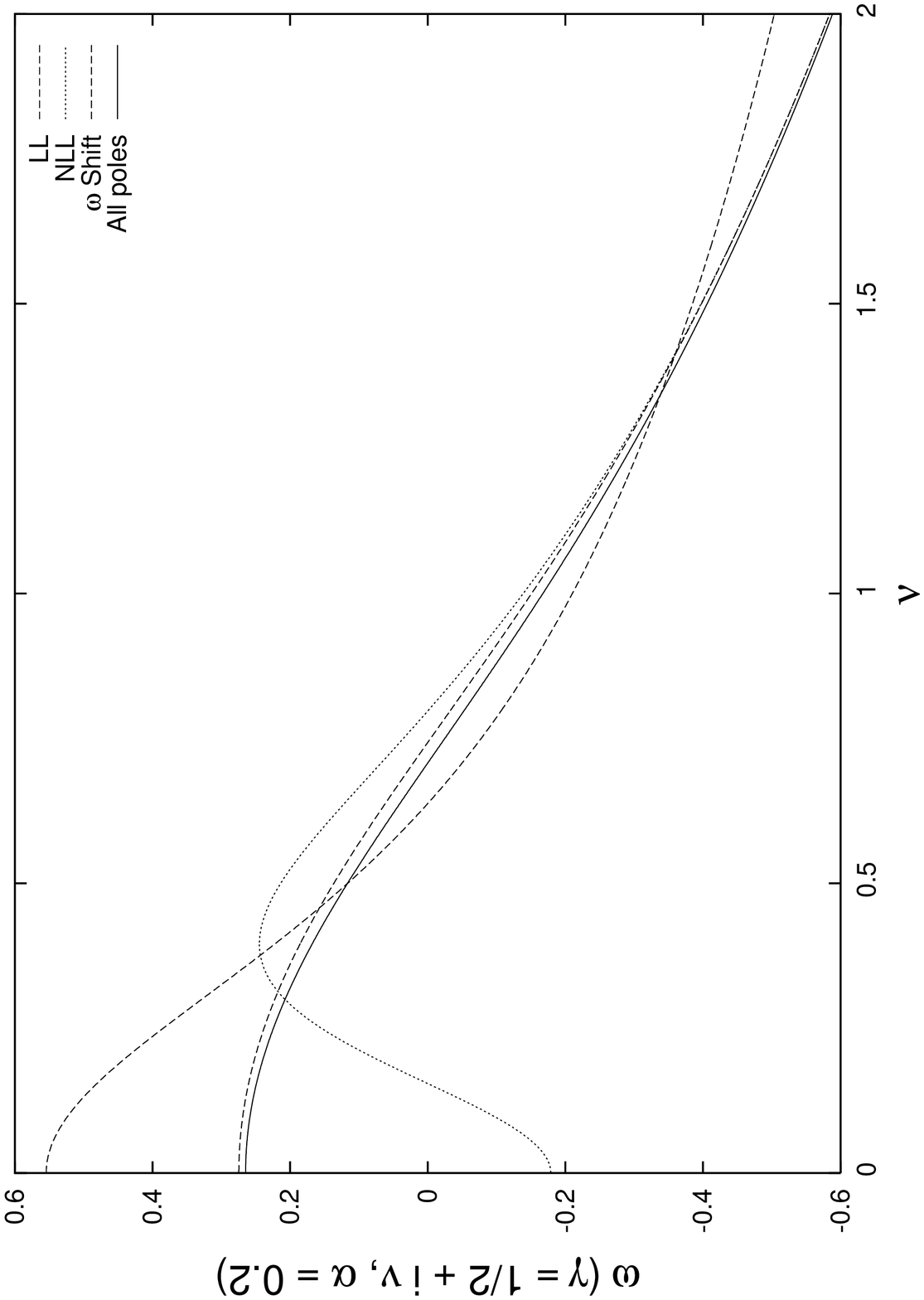}
\hspace{1cm}
\includegraphics{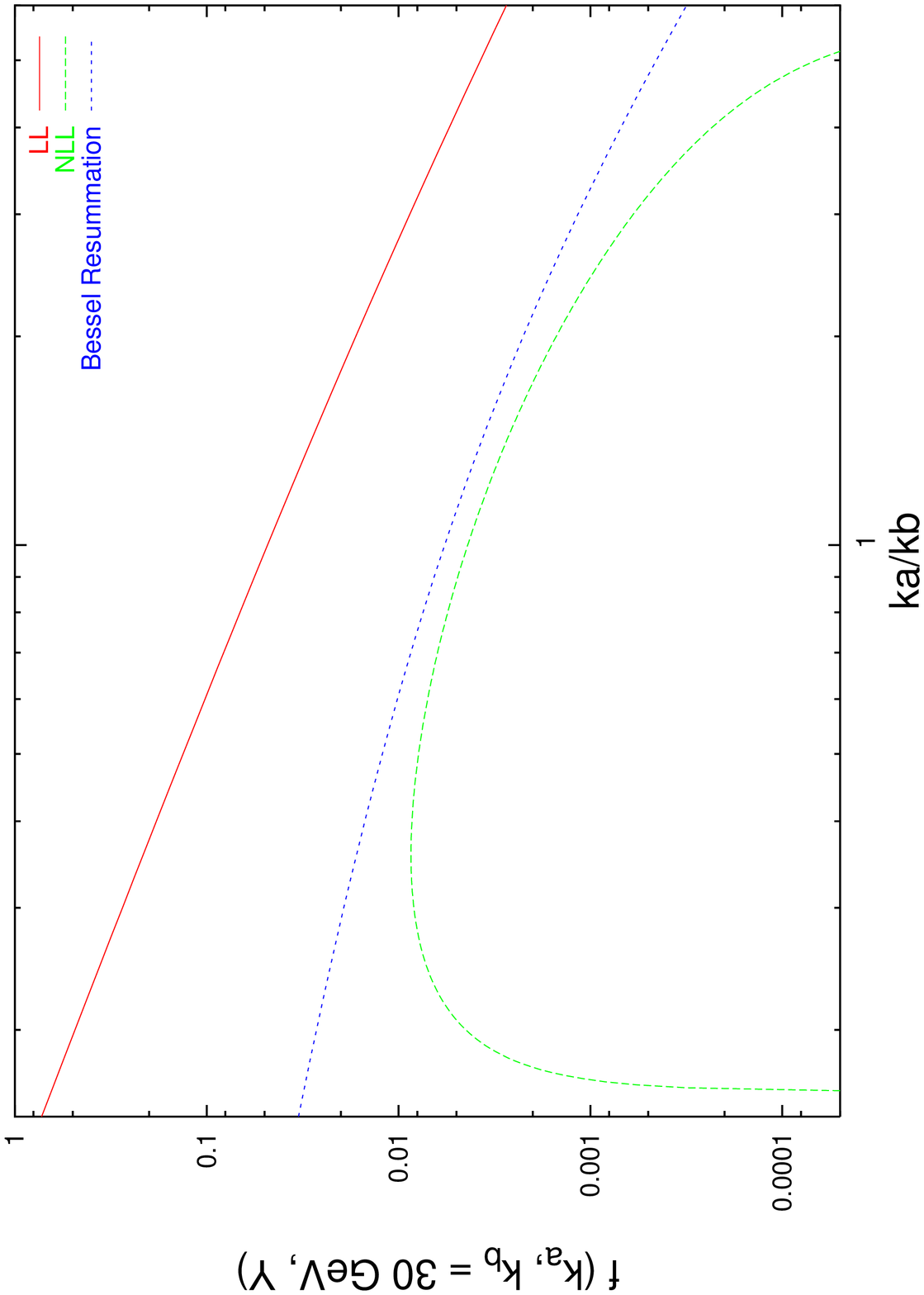}
  \caption{The LL and NLL scale 
invariant kernels together with the $RG$--improved kernel and the 
``all--poles'' resummation (left). The collinear behaviour of the NLL GGF 
using the Bessel resummation (right).}
\label{Agus1}
\end{figure}

The numerical solution to Eq.~(\ref{shiftGS}) and expression~(\ref{All-poles}) 
are compared in Fig.~\ref{Agus1} (left).
The expansion is now stable in all regions with an intercept of 0.3 at NLL for 
$\bar{\alpha}_s = 0.2$ (without running coupling effects). To implement 
Eq.~(\ref{All-poles}) in $k_t$--space is simple~\cite{Vera:2005jt}: Removing the 
term $-\frac{\bar{\alpha}_s^2}{4} \ln^2{\frac{q^2}{k^2}}$ 
from the real emission kernel, ${\cal K}_r \left(\vec{q},\vec{k}\right)$, 
and replacing it with
\begin{eqnarray}
\label{presc2}
\left(\frac{q^2}{k^2}\right)^{-{\rm b}{\bar \alpha}_s 
\frac{\left|k-q\right|}{k-q}}
\sqrt{\frac{2\left({\bar \alpha}_s+ {\rm a} \,{\bar \alpha}_s^2\right)}{\ln^2{\frac{q^2}{k^2}}}} 
J_1 \left(\sqrt{2\left({\bar \alpha}_s+ {\rm a} \,{\bar \alpha}_s^2\right) 
\ln^{2}{\frac{q^2}{k^2}}}\right) 
- {\bar \alpha}_s - {\rm a} \, {\bar \alpha}_s^2
+ {\rm b} \, {\bar \alpha}_s^2 \frac{\left|k-q\right|}{k-q}
\ln{\frac{q^2}{k^2}},
\end{eqnarray}
with $J_1$ the Bessel function of the first kind. This prescription generates a 
convergent GGF as can be seen in Fig.~\ref{Agus1} (right) 
where there are no oscillations, and can be immediately implemented in the 
iterative approach of~\cite{Andersen:2003an}. 

\begin{figure}
\centering
\vspace{5cm}  
\includegraphics{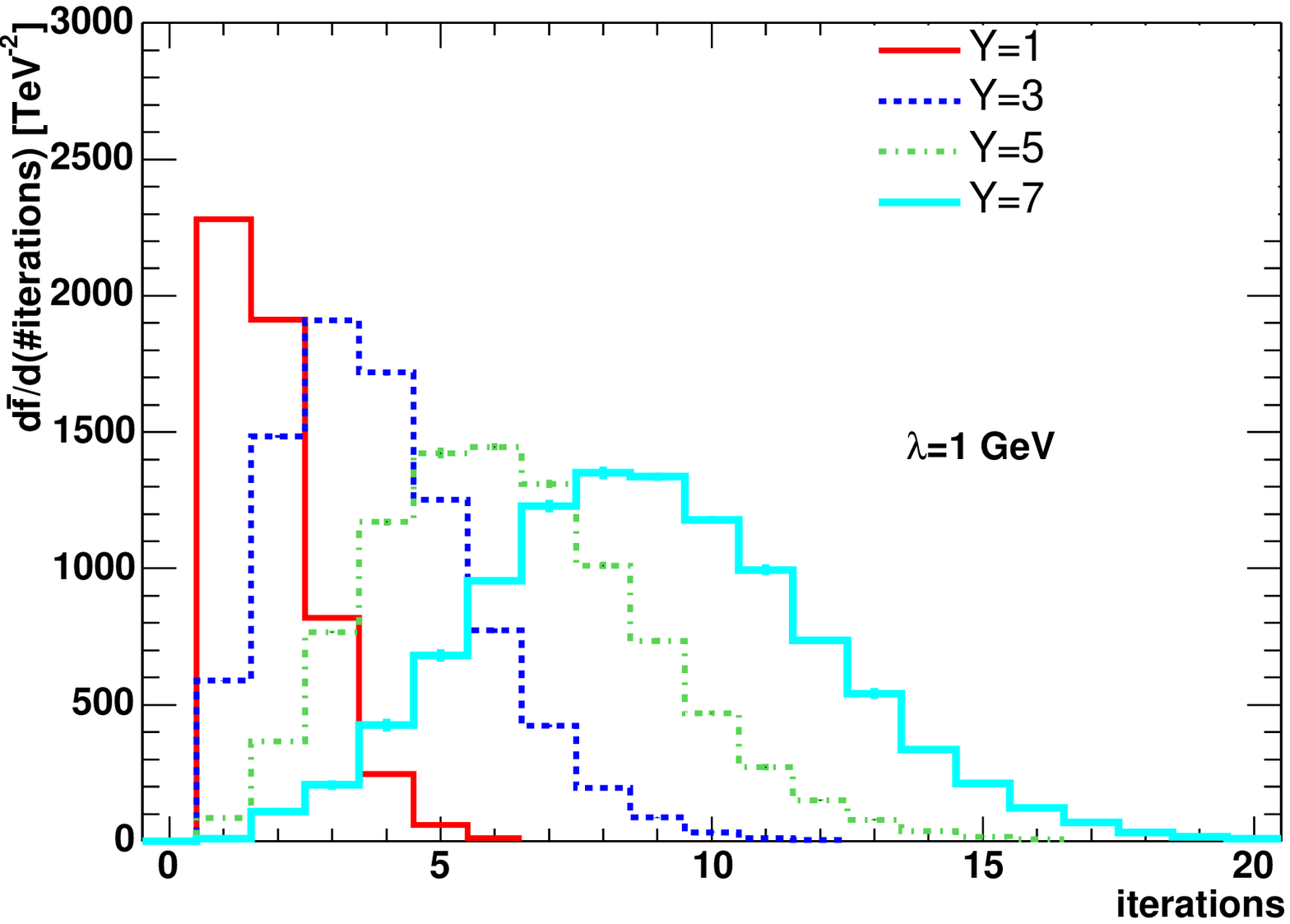}
\hspace{0.2cm}
\includegraphics{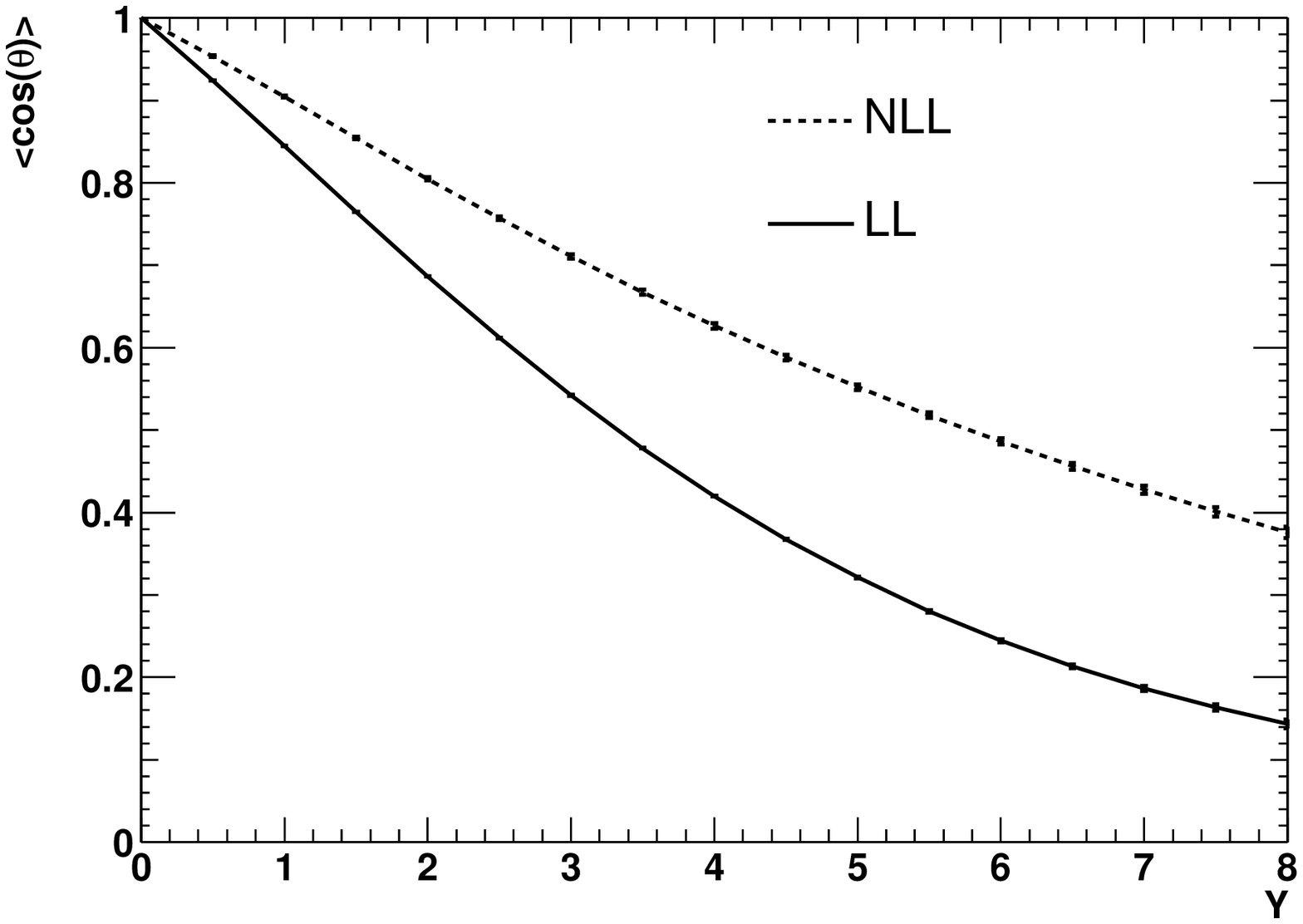}
\caption{Distribution in the number of iterations and angular dependence 
of the NLL gluon Green's function.}
\label{JAfigure2}
\end{figure}

This iterative method integrates the phase space using a
Monte Carlo sampling of different parton configurations. 
Multiplicities are extracted from the Poisson--like
distribution in the number of iterations needed to reach 
convergence (Fig.~\ref{JAfigure2} (left)).  Azimuthal angular correlations 
can also be obtained (Fig.~\ref{JAfigure2} (right))~\cite{Andersen:2003wy}.
To show how the angular dependences are correctly described the NLL kernel in 
$N=4$ super Yang--Mills theory~\cite{Kotikov:2000pm} 
was studied in~\cite{Andersen:2004uj}. In particular, 
to calculate the contribution to the GGF from its Fourier components 
in the azimuthal angle one can extract the coefficients either using the kernel 
for different conformal spins as in~\cite{Kotikov:2000pm}:
$\sim \int d \gamma 
\left({\bf{k}_a^2}/{\bf{k}_b^2}\right)^{\gamma}
\exp{(\omega_n (a,\gamma) {\rm Y})}$, 
or using the iterative solution~\cite{Andersen:2004uj}: 
$\sim \int_0^{2 \pi} {d\theta} \, 
f\left(\bf{k}_a,\bf{k}_b, {\rm Y}\right) \cos{\left(n \theta\right)}$.  
These two procedures match in their predictions and it can be seen that 
the $n=0$ Fourier component governs at large energies, decreasing the angular 
correlations.

The non--forward LL case was studied in~\cite{Andersen:2004tt} and the same 
method applies at NLL. In~\cite{Andersen:2004tt} it 
was shown how at large momentum transfer $t$ it is possible to 
study the diffusion into low and large scales of the transverse momenta in the 
gluon ladder. In particular, the diffusion into the infrared is cut off for 
finite values of $t$.   

\section*{Acknowledgements}
Thanks to J.~R.~Andersen for collaboration, and J.~Bartels, S.~Brodsky, V.~Fadin, 
L.~N.~Lipatov, R.~Peschanski, L.~Szymanowski, G.~P.~Vacca, H~de~Vega and 
S.~Wallon for fruitful discussions during this workshop. Support from the Alexander von Humboldt Foundation is acknowledged.


\section*{References}
\begin{thebibliography}{99}

\bibitem{FKL}  L.\thinspace N.~Lipatov, Sov.\ J.\ Nucl.\ Phys.\ {\bf 23}, 338 (1976); V.\thinspace S.~Fadin, E.\thinspace A.~Kuraev and L.\thinspace N.~Lipatov, Phys.\ Lett.\ B {\bf 60}, 50 (1975), Sov.\ Phys.\ JETP {\bf 44}, 443 (1976), Sov.\ Phys.\ JETP {\bf 45}, 199 (1977); I.\thinspace I.~Balitsky and L.\thinspace N.~Lipatov, Sov.\ J.\ Nucl.\ Phys.\ {\bf 28}, 822 (1978), JETP\ Lett.\ {\bf 30}, 355 (1979).

\bibitem{FLCC} V.\thinspace S.~Fadin and L.\thinspace N.~Lipatov, Phys.\ Lett.\ B {\bf 429}, 127 (1998); G.~Camici and M.~Ciafaloni, Phys.\ Lett.\ B {\bf 430}, 349 (1998).

\bibitem{Salam}
  G.~P.~Salam, JHEP {\bf 9807}, 019 (1998).

\bibitem{Ciafalonietal}
M.~Ciafaloni, D.~Colferai, G.~P.~Salam and A.~M.~Stasto, 
Phys.\ Lett.\ B {\bf 587} (2004) 87, Phys.\ Rev.\ D {\bf 68} (2003) 114003, 
Phys.\ Lett.\ B {\bf 576} (2003) 143, Phys.\ Lett.\ B {\bf 541} (2002) 314, 
Phys.\ Rev.\ D {\bf 66} (2002) 054014; 
M.~Ciafaloni, D.~Colferai and G.~P.~Salam, JHEP {\bf 0007} (2000) 054, 
JHEP {\bf 9910} (1999) 017, Phys.\ Rev.\ D {\bf 60} (1999) 114036; 
M.~Ciafaloni and D.~Colferai, Phys.\ Lett.\ B {\bf 452} (1999) 372.

\bibitem{NLOpapers} 
L.N. Lipatov, {JETP}  ${\bf {63}}$, 904 (1986); 
G. Camici and M. Ciafaloni,  {Phys. Lett.} B  ${\bf {395}}$, 118  (1997); 
D.~A.~Ross, Phys.\ Lett.\ B {\bf 431} (1998) 161; 
J.~R.~Forshaw, D.~A.~Ross and A.~Sabio Vera, Phys.\ Lett.\ B {\bf 455} (1999) 273, 
Phys.\ Lett.\ B {\bf 498} (2001) 149; 
M.~Ciafaloni, M.~Taiuti and A.~H.~Mueller, Nucl.\ Phys.\ B {\bf 616} (2001) 349;
Yu.V.~Kovchegov and A.H.~Mueller, Phys. Lett. {\bf B439} (1998) 423; 
N.~Armesto, J.~Bartels, M.A.~Braun, Phys. Lett. {\bf B442} (1998) 459; 
S.~J.~Brodsky, V.~S.~Fadin, V.~T.~Kim, L.~N.~Lipatov and G.~B.~Pivovarov, JETP Lett.\  {\bf 70} (1999) 155; 
C.~R.~Schmidt, Phys.\ Rev.\ D {\bf 60} (1999) 074003; 
G.~Chachamis, M.~Lublinsky and A.~Sabio~Vera, Nucl.\ Phys.\ A {\bf 748} (2005) 649;
G.~Altarelli, R.~D.~Ball and S.~Forte, Nucl.\ Phys.\ B {\bf 674} (2003) 459, 
Nucl.\ Phys.\ B {\bf 621} (2002) 359, Nucl.\ Phys.\ B {\bf 599} (2001) 383, 
Nucl.\ Phys.\ B {\bf 575} (2000) 313.

\bibitem{Andersen:2003an}
  J.~R.~Andersen and A.~Sabio Vera, Phys.\ Lett.\ B {\bf 567} (2003) 116.

\bibitem{Andersen:2003wy}
  J.~R.~Andersen and A.~Sabio Vera, Nucl.\ Phys.\ B {\bf 679} (2004) 345.

\bibitem{Vera:2005jt}
  A.~Sabio~Vera,  Nucl.\ Phys.\ B {\bf 722} (2005) 65.

\bibitem{Kotikov:2000pm}
  A.~V.~Kotikov and L.~N.~Lipatov,  Nucl.\ Phys.\ B {\bf 582} (2000) 19.

\bibitem{Andersen:2004uj}
  J.~R.~Andersen and A.~Sabio Vera, Nucl.\ Phys.\ B {\bf 699} (2004) 90.

\bibitem{Andersen:2004tt}
  J.~R.~Andersen and A.~Sabio~Vera, JHEP {\bf 0501} (2005) 045.

\end{thebibliography}
\end{document}